\documentclass[reprint,amsmath,amssymb,aps,longbibliography]{revtex4-2}
\usepackage[T1]{fontenc}
\usepackage{graphicx}
\usepackage{dcolumn}
\usepackage{bm}
\usepackage{amsmath}
\usepackage{lmodern}
\usepackage{mathtools}
\usepackage{siunitx}
\usepackage{multirow}
\usepackage[colorlinks,citecolor=blue,linkcolor=blue,urlcolor=blue]{hyperref}
\usepackage[normalem]{ulem}

\def\ee{{\rm e}}

\begin{document}

\title{The role of all-optical neural networks}
\author{M.~Matuszewski$^1$}
\author{A.~Prystupiuk$^{1,2}$}
\author{A.~Opala$^{1,2}$}

\affiliation{$^1$The Institute of Physics, Polish Academy of Sciences, Aleja Lotnik\'ow 32/46, PL-02-668 Warsaw, Poland}
\affiliation{$^2$Institute of Experimental Physics, Faculty of Physics, University of Warsaw, ul. Pasteura 5, PL-02-093 Warsaw, Poland}

\begin{abstract}
In light of recent achievements in optical computing and machine learning, we consider the conditions under which all-optical computing may surpass electronic and optoelectronic computing in terms of energy efficiency and scalability. When considering the performance of a system as a whole, the cost of memory access and data acquisition is likely to be one of the main efficiency bottlenecks not only for electronic, but also for optoelectronic and all-optical devices. However, we predict that all-optical devices will be at an advantage in the case of inference in large neural network models, and the advantage will be particularly large in the case of generative models. We also consider the limitations of all-optical neural networks including footprint, strength of nonlinearity, optical signal degradation, limited precision of computations, and quantum noise. 
\end{abstract}
\maketitle

\section{\label{sec:intro}Introduction} 

In recent years, remarkable strides have been made in the field of machine learning and artificial intelligence, heralding a new era of practical applications that are swiftly permeating various industries and our daily lives. However, this remarkable progress comes at the price of the rise in energy consumption, which is primarily driven by the exponential growth in the volume of data being processed~\cite{mehonic2022brain} and to the apparent flattening of the improvement of computing performance. For many decades, progress was governed by remarkable principles of Moore's law and Dennard scaling. However, it seems that we are now entering a phase where these principles are gradually approaching a more stable plateau~\cite{Waldrop_Moore,Xu_EdgeInference}. It has become apparent that the cost of data movement through electronic wires, requiring charging their capacity for each bit of information, dominates the energy budget for data-intensive applications such as large-scale machine learning~\cite{verma2019memory,Miller_Attojoule}.

To circumvent this limitation and make computations more efficient, a natural strategy is to reduce the physical distance between memory and processing units. This motivates interest in computing systems that go beyond the von Neumann architecture, such as in-memory computing~\cite{verma2019memory}. Many physical implementations of machine learning have been realized with emulators of neural networks on specialized hardware, where the structure of the network is resembled physically~\cite{Misra_ANNSurvey,Merolla,Loihi,Qian_MemristorCNN}. In these cases, computing is often analog rather than digital. Since neural network models are themselves analog, this approach appears to be more adequate in the case of neural networks than in the case of traditional algorithm-based computing.

Another way to increase the efficiency of computations is to realize them on a non-standard physical platform~\cite{Grollier_review,Huang_ReviewNeuromorphic,Tanaka}. A particularly promising approach is to use photons instead of electrons~\cite{Tait_LightTech_2014,Shastri_review,Wetzstein_review,Tanaka,Ballarini_Neuromorphic,Opala_NeuromorphicComputing,Englund2019,stroev2023renaissance}. The advantage of optical systems is that they do not require charging the capacitance of communication channels, so data movement can be almost lossless. For this reason, optical systems are used for communications over large distances and at high data rates, when the energy cost of data movement is particularly important~\cite{Agrell_2016}. However, computation with photons, while being researched for many decades~\cite{goodman1978fully}, has not found mainstream applications yet. Optical computing has been hampered by many factors including the weakness of optical nonlinearity, bulkiness of optical elements, difficulty to regenerate optical signals and to integrate optical sources. It has been difficult to realize a device capable of realizing general digital operations with appropriate fidelity and at a low energy cost~\cite{Miller_OpticalLogic}. However, the recent resurgence of interest in optical computing led to breakthroughs that alleviated many of these limitations~\cite{Shastri_review,Wetzstein_review}, and it seems that taking advantage of optical computing in practical devices is within reach.

A natural solution to overcome the disadvantages of electronic and optical computing is to combine them in a single system, using the advantages of both light and matter. This approach typically assumes constructing an optoelectronic device where communication or linear operations are realized optically, while other operations, including nonlinear transformations and signal regeneration, fan-in, and fan-out are taken care of by electronics. While this approach is very promising, it has its own limitations. One of them is the limited compatibility of electronic and optical systems and the difficulty to integrate the two. For example, typical length scales for state-of-the-art electronics are in the range of a few nanometers. In the case of optics, it is difficult to squeeze light below the micrometer length scale without incurring significant losses. On the other hand, conversion from electronic to optical signals and vice versa and from analog to digital signals creates additional energy costs and technological difficulties.

Here, we consider the viability of all-optical computing as an alternative to electronic and optoelectronic approaches, and attempt to identify the applications where it may excel. This topic has been considered previously, and it was pointed out that an all-optical approach may not be well suited for digital computations~\cite{Miller_OpticalLogic}. We look at this problem from a new perspective, taking into account recent advancements both in the optical technology and in the field of machine learning. We assume that the main technological bottleneck of high-intensity computing is, as it appears, the efficiency of data movement via electronic wires. This is justified  both by fundamental physical limitations, and by the observation that electronic computation efficiency is apparently saturating after decades of exponential progress. On the other hand, there is still room for improvement before the fundamental limits are reached for optics. Consequently, we consider the energy cost of operations requiring data movement by electronic channels, such as memory access, to be the most stringent limitation. Importantly, to provide a fair comparison, we consider the efficiency of the complete computing system, and not only a certain part of it. In particular, we take into account the cost of data acquisition, and the electronic memory access cost necessary to provide data, which is often overlooked in estimates of energy efficiency. It is important to mention that our considerations do not apply to the case where input data can be supplied in the form of optical signals, without accessing external  electronic memory~\cite{ashtiani2022chip,wang2023image}.

Based on these assumptions, we try to answer the question  of the viability and practicality of all-optical neural networks. In other words, we consider whether there are applications in which all-optical neural networks can outperform their electronic and optoelectronic counterparts, and to what extent. We conclude that electronic memory access cost will be likely the main limitation not only for electronic and optoelectronic, but also for all-optical networks. However, we find that one application where all-optical networks will be at an advantage is  the inference in large-scale neural networks, where the number of neurons in the hidden layer is much larger than the dimensionality of inputs and outputs. The advantage will be particularly large in the case of generative models, where input data can be reused in many subsequent inferences, reducing data acquisition costs. These conditions are fulfilled in many machine learning models used in practice. 

In addition, we consider the limitations of all-optical neural networks that must be overcome before they become practical. We analyze optical neural networks  taking into account the specifics of information processing with light, such as quantization of light.  By performing numerical simulations, we show that all-optical neural networks can be accurate even if the precision of optical transformations is reduced by noise and fabrication errors. We discuss the issues of signal regeneration, network depth, and scalability of optical networks.

\section{\label{sec:alloptical} Can all-optical neural networks be efficient?} 

In this section, we show the main motivation of our paper, that is the advantage of using all-optical systems as an efficient platform for analog neural networks, as opposed to electronic or optoelectronic devices. We consider the energy cost of calculations, which is currently the most important limitation of computing systems~\cite{Waldrop_Moore,verma2019memory}. We leave the considerations of footprint, speed, precision, and other limitations of optical systems to the next section. 

The main assumption of our considerations is that the data movement cost in electronic wires will be difficult to improve in the future. This assumption can be justified by two arguments. One is the physical lower limit of the energy required to charge an electronic wire to send a single bit of information. The cost of charging wire capacitance per unit length is approximately independent of the wire cross section, and can be estimated as 100 fJ/bit per 1 mm of connection length~\cite{Miller_Attojoule}. Another argument for considering data movement cost as a physical lower limit is the apparent saturation of energy efficiency of computations and memory access costs~\cite{Xu_EdgeInference,Waldrop_Moore,verma2019memory}, despite decades-long developments and huge investments in the complementary metal-oxide (CMOS) technology. In fact, it appears that state-of-the-art efficiencies are reaching the fundamental estimates. In machine learning applications, which require great number of memory access operations to perform multiplications of large tensors, the cost of data movement is at least comparable the cost of logic operations~\cite{verma2019memory}. Therefore, it seems that the room for improvement for the efficiency of current CMOS technology is limited, unless a significant technological breakthrough is achieved.

How optics can be advantageous from the perspective of hardware-implemented neural networks? As was mentioned, optical links do not require charging of communication lines. Optical energy dissipation corresponds to effects such as optical absorption, light leakage in waveguides, optical dispersion, and spatial or temporal mode misalignment, however these effects typically lead to a much lower dissipation than in the case of electronics. This is the reason why optical connections are used for long-haul communications, in data centers, and can be applied for communications even at short length scales~\cite{Miller_OpticalInterconnects}.

In this work, we focus on the aspects specific to neural networks. The structure of neural networks and the specifics of the required computations make optics a much better match than in the case of algorithmic digital computations. In this context, one can point out several advantages that we list below.

\subsection{Optical fan-out and fan-in} 

In a typical artificial neural network, neurons perform two kinds of operations. Summation of neuron inputs $x_i$ multiplied by weights $w_{ij}$ is a linear operation (i.e.~linear as a function of neuron inputs), which is followed by a nonlinear activation given by a function $f_j$ such as the sigmoid or rectified linear unit (ReLU)
\begin{equation} \label{eq:neuron}
    y_j=f_j\left(\sum_i^N w_{ij} x_i\right).
\end{equation}
Note that $f_j$ can act on vectors rather than scalars as is the case in the softmax function. In the case of electronic systems, the efficiency is strongly tied to the cost of a single multiply and accumulate operation (MAC). This operation occurs once for every multiplication of a neuron input $x_i$ with the corresponding input (synaptic) weight $w_{ij}$. Performing such an operation in the von Neumann machine requires accessing memory for all the inputs and all the corresponding weights. If the number of neuron inputs is large, so is the required data movement. On the other hand, the nonlinear activation function is applied only once per neuron activation, and its energy cost can be much lower. In practice, input-weight multiplications are performed in batches, so the weights can be to a large extent reused if stored in a local memory. However, due to the scale of large machine learning models, the memory access cost is still a major source of energy dissipation. 

On the other hand, in the case of optical neural networks, the summation of optical signals can be performed at almost no cost of data movement by directing or focusing optical pulses or beams to certain regions in space, which is the optical fan-in. This can take the form of either simple intensity addition in the case of mutually incoherent light pulses, or optical interference in the case of mutually coherent pulses. On the other hand, fan-out of output optical signals to a very large number of copies can be realized with linear optical elements such as diffractive optical elements~\cite{bernstein2022single}, spatial light modulators, microlens arrays~\cite{wang2023image}, or in integrated circuits~\cite{Soljacic_DeepLearning}. There is no fundamental lower limit for the energy cost of these operations. Therefore, it is the nonlinear activation function, rather than the weighted linear summation that creates a bottleneck for the highest possible efficiency of all-optical devices. This is a particularly important limitation due to the weakness of nonlinear interactions between photons, which are much weaker than the interactions between electrons in semiconductors.

\subsection{Static weights in neural network inference} 

In machine learning, the inference phase follows the training phase. From the point of view of energy consumption, the inference phase is often more important that the training phase, since the trained model can be used for inference arbitrarily many times~\cite{Jouppi_InDatacenter}. Specialized CMOS inference systems include Google TPUv1 and nVidia inference platform. In the inference phase, weights of neurons do not change. Therefore, if weights can be implemented in optical hardware without the need for external memory access, the cost of data movement can be greatly reduced. In the case of electronics, this approach is the basis of in-memory computing~\cite{verma2019memory}. However, electronic chips with in-memory computing can require complicated wiring to connect computing units with each other~\cite{Merolla}. In the case of optics, hardware-encoded static weights can be combined with almost dissipationless transport to drastically reduce the cost of weighted summation in Eq.~(\ref{eq:neuron}).
The multiplication of inputs by the corresponding weights can be implemented with linear optical elements which apply a certain amplitude or phase modulation to the optical signals. One of widely used methods is implementing Mach-Zehnder interferometers, or a mesh of such interferometers that perform an arbitrary linear operation represented by a matrix~\cite{Soljacic_DeepLearning}. Weights encoded in phase change materials~\cite{Feldmann_AllOpticalSpikingNetwork} do not require power to sustain their state. In the case of free-space propagating beams, spatial light modulators can be used for applying weights, with millions of independently tunable parameters~\cite{bernstein2022single}. While some of these methods require an energy supply to keep the state of optical weights, this cost can be often reduced to a level that is much below the energy dissipation cost of data movement in electronics. In some other cases, it can be completely eliminated.

\subsection{Structure of large neural network models} 

One of the reasons for the recent progress in machine learning is that hardware, such as specialized parallel computing units, allowed the implementation of models with increasing complexity, that could accommodate and process large datasets. Usually, to obtain high accuracy of predictions, it is necessary to use models with a large number of parameters and neurons. It was noticed that optics could be particularly effective in comparison to electronics in applications that require large scale computations~\cite{nahmias2019photonic}. 
\begin{figure}
    \centering
    \includegraphics[width=0.8\columnwidth]{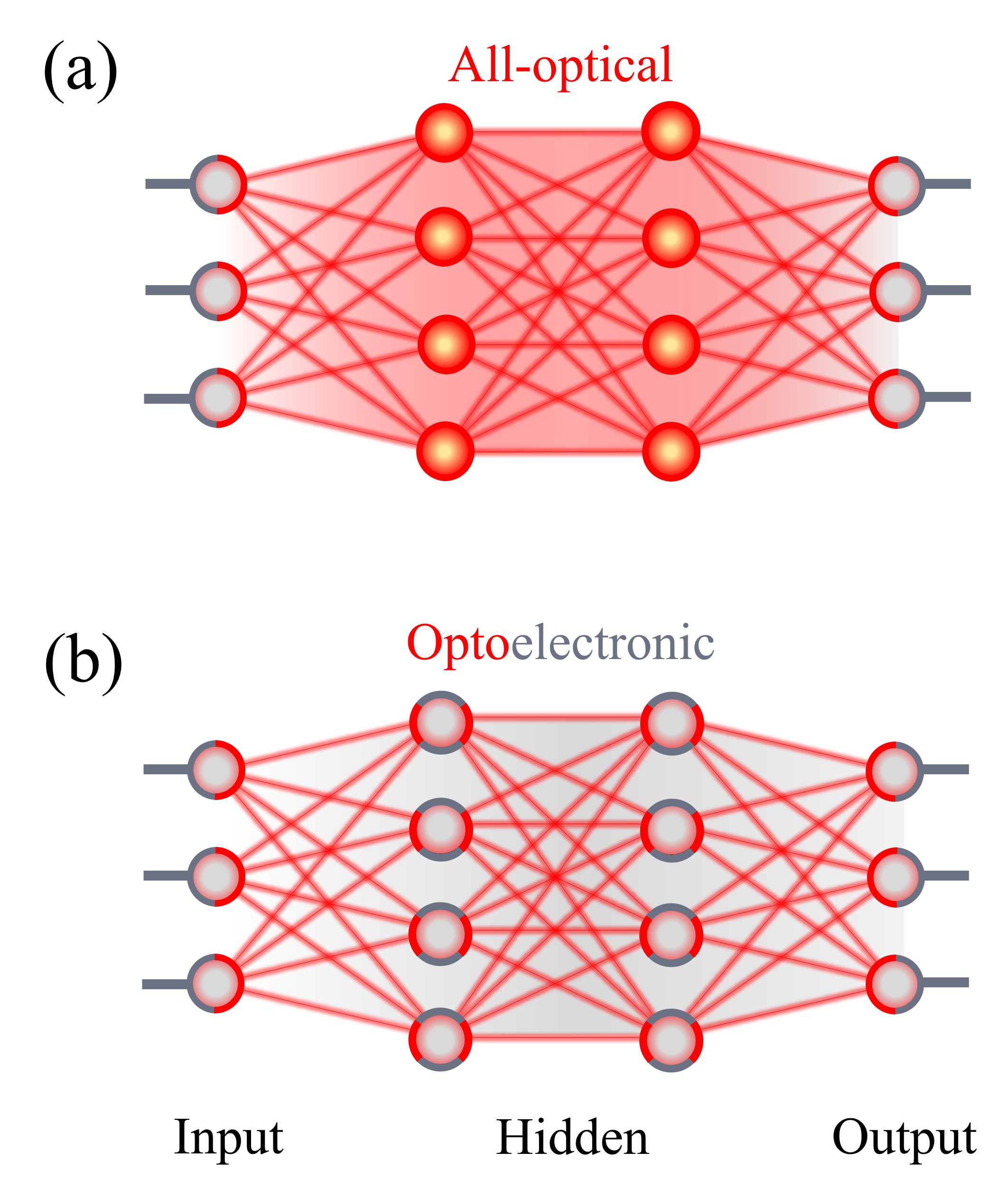}
    \caption{All-optical and optoelectronic neural networks. (a) In an all-optical network, input data is transformed to optical form at the input layer, and all subsequent operations up to the output layer are realized all-optically. (b) In an optoelectronic network, the signal is transformed from optical to electronic and back at each layer of the network. If the size of the hidden (middle) layers is much larger than input and output layers, this results in a bottleneck of system efficiency.}
    \label{fig:alloptical}
\end{figure}
This advantage is particularly large in the case of all-optical networks. Let us, for the moment, focus on the memory access cost as the main bottleneck and compare the potential efficiency of all-optical and optotelectronic devices (we will justify this assumption later). In the case of a typical optoelectronic device, the data needs to be converted to a digital signal and stored in memory after each layer of neural network computation, while in the all-optical system this is not necessary, see Fig.~\ref{fig:alloptical}. We assume that the input to an all-optical device is provided in an electronic form since most of the data in our world is encoded electronically. It needs to be read from a digital memory both in the case of an optoelectronic and an all-optical neural network. However, the advantage of an all-optical network is that once converted to the optical domain, it does not need to be re-converted to the electronic, digital domain until the entire computation on the data sample is finished. In Fig.~\ref{fig:alloptical}, memory access cost occurs only at the input and the output of the all-optical network, and at each hidden neuron of an optoelectronic network. Moreover, in the cases where data is provided in an optical form, it does not occur at the input layer~\cite{ashtiani2022chip,wang2023image}. 

\subsection{Comparison of optotelectronic and all-optical network efficiencies} 

How does this lower memory access cost affect the efficiency of optical networks? This largely depends on the particular structure of the neural network model, and we analyze some examples here. Generally, in the case of large scale neural networks, the size of hidden layers in Fig.~\ref{fig:alloptical}, measured as the number of neurons or the number of parameters, is much larger than the size of input and output layers~\cite{vaswani2017attention,Amoeba}. The energy cost of computations per sample in the inference stage can be very roughly estimated for an optoelectronic network as
\begin{equation}\label{eq:OE}
 E_{\rm OE} \geq E_{\rm electronic} (N_{\rm input} + N_{\rm output} + N_{\rm hidden}),
 \end{equation}
and for an all-optical network as
 \begin{equation}\label{eq:AO}
 E_{\rm AO} \geq E_{\rm electronic} (N_{\rm input} + N_{\rm output}) + E_{\rm optical} N_{\rm hidden},
\end{equation}
where $E_{\rm electronic}$ is the average cost of electronic operations per neuron per inference, including memory access cost, and $E_{\rm optical}$ is the average cost of optical operations per neuron per inference, including the energy of light pulses, optical losses, and electronics necessary for an optical neuron to operate. Under the assumption that the memory access cost is the main bottleneck of computation efficiency, and considering the case where the majority of neurons are in the  hidden layer, $N_{\rm hidden}\gg N_{\rm input} + N_{\rm output}$, the ratio of energy costs can be estimated as
\begin{align*}
    \frac{E_{\rm AO}}{E_{\rm OE}} &= \frac{N_{\rm input} + N_{\rm output}}{N_{\rm input} + N_{\rm output} + N_{\rm hidden}} +\\ &+ \frac{E_{\rm optical}}{E_{\rm electronic}}\frac{N_{\rm hidden}}{N_{\rm input} + N_{\rm output} + N_{\rm hidden}} \approx \\
    &\approx \frac{E_{\rm optical}}{E_{\rm electronic}}
\end{align*}
If $E_{\rm electronic} \gg E_{\rm optical}$, i.~e.~the energy cost of computation per inference per neuron in the optoelectronic network is much larger than the cost of the same operation performed all-optically, the energy cost will be much lower in the case of an all-optical device. 

To justify the above reasoning, we consider whether the above conditions: $N_{\rm hidden}\gg N_{\rm input} + N_{\rm output}$ and $E_{\rm electronic} \gg E_{\rm optical}$, can be fulfilled in practice. 
To  estimate the average cost of electronic operations  $E_{\rm electronic}$ in an optoelectronic device, one needs to take into account the costs of conversion from an optical signal to electronic signal, analog to digital conversion, the costs of the reverse processes, and the cost of memory access. In certain optoelectronic devices, some of these costs may be absent, for example, if the electronic part of computation is analog as well. However, it appears to be difficult to reduce the cost of all these operations below the level of picojoules per data unit such as a byte. In particular, the cost of memory access appears to be the main bottleneck. For an 8-bit input, it ranges from several  picojoules to several nanojoules depending on the technology used and the size of memory. For example, single access to 100 MB memory requires around 10 pJ of energy~\cite{verma2019memory,Xu_EdgeInference}.  Even if the system is designed in such a way that access to memory is not required for each neuron operation, the cost of analog-to-digitial conversion and electronic to optical conversion results in a similar bottleneck~\cite{tait2022quantifying,chen2022deep,Adept}. 

On the other hand, all-optical neural networks in the inference mode do not require optoelectronic conversion and the energy $E_{\rm optical}$ is mainly bounded by the required power of the light source. This bound depends on the optical nonlinearity of the system, optical losses, and the sensitivity of detectors at the output layer. While weak nonlinear response is one of the main disadvantages of optical systems, the use of materials characterized by strong and ultrafast nonlinearities, such as semiconductor quantum wells, organic materials or two-dimensional materials can result in high efficiency of nonlinear operations at high data rates. In particular, exciton-polaritons are quasiparticles of light and matter that can exhibit optical nonlinearity that is orders of magnitude stronger than in other materials. Using exciton-polaritons, the energy cost of a single nonlinear operation can be as low as few attojoules per neuron~\cite{Matuszewski_EnergyEfficient}. Taking into account the possibility of optical fan-in with thousands of linear operations per neuron~\cite{Englund2019},  optical efficiency at the level of zeptojoules per operation is foreseeable. At the same time, the required sensitivity of detectors scales proportionally to $N_{\rm hidden}/N_{\rm output}$, since  at the output layer the light is collected from all of the hidden neurons. Accordingly, in the limit of large $N_{\rm hidden}/N_{\rm output}$ considered here, detector sensitivity will not be the main bottleneck. 

Large size of a neural network model translates to a large size of hidden layers $N_{\rm hidden}$. As a result, large-scale neural networks used in practice are usually characterized by very high ratios $N_{\rm hidden}/N_{\rm input}$. For example, one of the leading models in the ImageNet competition, Amoeba-Net, has $10^9$ hidden nodes and performs $10^{11}$ operations per inference, while the ImageNet input size is $165\times 165 \times 3 \approx 10^5$, resulting in $N_{\rm hidden}/N_{\rm input}\approx 10^4$. In recent language models, this ratio is even higher, with the large BERT model consisting approximately of $24\times2\times512\times 1024$ nonlinear nodes for a $512$-long input token sequence, resulting in the ratio $N_{\rm hidden}/N_{\rm input}\approx 10^5$. Therefore, the condition $N_{\rm hidden}\gg N_{\rm input} + N_{\rm output}$ is fulfilled in many practical large neural network models. It is interesting that the same relation appears to hold for the network of neurons in the human brain. The number of neurons in the brain is of the order of $10^{11}$, which is likely to be much greater than the  dimensionality of the input information from all stimuli.

To give a concrete example of potential efficiency, we estimate the energy cost per operation for a hypothetical large-scale neural network with $N_{\rm input}+N_{\rm output}=10^3$ and $N_{\rm hidden}=10^8$, assuming $E_{\rm electronic}=1\,$pJ, and $E_{\rm optical}=100\,$aJ. In both cases, we assume the optical fan-in of 1000 inputs per neuron in the hidden layer. For a fair comparison, the energy cost per operation is calculated as the total energy cost of the complete network, including memory access for each electronic neuron operations. The number of operations is calculated as two operations (multiply and accumulate) per each neuron input and one for nonlinear activation in a neuron. According to Eqs.~(\ref{eq:OE}) and (\ref{eq:AO}) the lower bounds for the energy cost are estimated to be 500 aJ for an optoelectronic network and 55 zJ for an all-optical network, almost four orders of magnitude lower. 

These estimations are not complete unless we consider the cost of acquiring data. This includes the cost of access to external memory, such as DRAM memory, from where input data has to be retrieved. In the case of data that needs to be transmitted over a distance, as is usually the case in cloud computing, the cost of accessing input data may be further increased. The costs of both reading from DRAM memory and fiber link data transmission are in the range of 1-100 pJ per bit, or up to 1 nJ per byte~\cite{verma2019memory,Thraskias_Interconnects}. These costs may be significantly higher in the case of wireless communication or less efficient data transmission channels. The  cost of acquiring input data may also be high in the case of edge computing, for example, if input information is gathered by a camera with a high energy cost per pixel. In all of these cases, the input data costs may dominate over all other costs of computations.

However, there is an important class of practical machine learning tasks where input data costs can be drastically reduced by ``recycling'' input data acquisition. All generative tasks in which most of the input information used at one step of computation can be used for the next step belong to this category. These include applications such as text completion, language translation, question answering, chatbots, image and sound synthesis. In these cases, input information known as "context" can be often reused to a great extent across inferences, for example, $4\times 10^3$ times in the case of large language models. As a result, the cost of data acquisition may be orders of magnitude smaller than the cost of local memory access for input neurons, which is already included in our estimations.

\begin{table}
\begin{tabular*}{\columnwidth}{@{\extracolsep{\fill} } l l l l } 
 \hline
 Network type & Small & Large & Large \\ &&&generative \\
 \hline \hline
 $N_{\rm input} + N_{\rm output}$ & 100 & 1000 & 1000\\
 $N_{\rm hidden}$ & 1000 & $10^8$ & $10^8$\\
 \hline
Electronic & 1 pJ & 1 pJ & 1 pJ \\ 
 Optoelectronic & 7 fJ & 1 fJ & 1 fJ \\ 
 All-optical & 6 fJ & 600 zJ & 100 zJ \\ 
 \hline
\end{tabular*}
\caption{Estimates of average energy cost per operation in inference for the system as a whole, including data acquisition cost. Parameters are $E_{\rm electronic}=2\,$pJ, $E_{\rm optical}=100\,$aJ, $E_{\rm memory}=10\,$pJ, $E_{\rm acquisition}=100\,$pJ. We assume $10^3$ inputs per neuron on average and the possibility to reuse input data $10^3$ times in a generative network. 
}
\label{table}
\end{table}

In Table~\ref{table} we present examples of complete estimates for electronic, optoelectronic, and all-optical neural networks in the case of various machine learning model sizes. We take into account the costs of all contributions to energy usage, including optical, optoelectronic, memory, and data acquisition. To this end, in the calculation of energy cost per inference we include additional terms in addition to the ones present in Eqs.~(\ref{eq:OE}) and~(\ref{eq:AO})
\begin{equation}
   E^{\rm total}_{\rm OE,AO} = E_{\rm OE,AO} + N_{\rm input} \left(\frac{E_{\rm acquisition}}{M} + E_{\rm memory}\right)
\end{equation}
where $E_{\rm acquisition}$ is the cost of acquiring input data, which is divided by the number of inferences $M$ where it is reused, and $E_{\rm memory}$ is the cost of accessing local memory that stores the input data. 
It is clear from Table~\ref{table} that all-optical neural networks will have the advantage in the case of large models, and in particular in the case of generative models which require less input data.

\section{\label{sec:limitations} Limitations of all-optical computing} 

In this section, we discuss the possible limitations of all-optical computing. We consider the footprint and speed of operations, cascadability and signal degradation, implementing useful nonlinear transformations, precision of computations, fabrication errors, and quantum noise.

\subsection{Footprint} 

One of the  arguments commonly raised against using optics for computing is the footprint of optical systems. The rationale of this argument is that the wavelength of visible light is of the order of a single micrometer, while electronic systems can be integrated in chips with nanometer-sized transistors. While the ffotprint is certainly a limitation for optics, it is actually not as severe as it may seem. In the case of neural network implementations in electronics, the nanometer size of a transistor does not directly translate into nanometer-sized neurons. For example, in the IBM TrueNorth neuromorphic chip~\cite{Merolla}, fabricated in 28-nm process technology, the footprint is approximately 200 $\mu$m$^2$ per neuron and 1 $\mu$m$^2$ per synaptic weight. Such length scales result from the complicated circuitry that must be implemented in an electronic chip to emulate a neuron. 

Moreover, it is important to realize that there exists a direct relation between the energy efficiency of a chip and its footprint, which results from the need to dissipate heat generated by the computation. Heat removal is space consuming. In CMOS chips, circuit structure is usually two-dimensional, and the third dimension is sacrificed for a heat sink. One exception to this rule are memory chips, which often have a multilayer stacked structure with more than 100 layers. This is possible due to the reduced amount of heat generation as compared to information processing chips. As a result, the reduction of energy dissipation leads to the reduction of footprint.

Moreover, there have been great advancements in the miniaturization of optical systems. Integrated silicon photonics chips can be fabricated and processed in large quantities by specialized foundries. A typical size of an element of a photonic chip, such as a Mach-Zehnder interferometer is of the order of micrometers~\cite{Soljacic_DeepLearning,Feldmann_AllOpticalSpikingNetwork,tait2019silicon}. If energy dissipation in such optical chips can is lower than the dissipation in electronic chips, stacking of optical chip layers should be possible, thus reducing the footprint.  On the other hand, the free-space approach to computing, while requiring the third dimension for light propagation, also permits achieving very low footprint. For example, commercially available spatial light modulators with a few $\mu$m pixel pitches are able to encode synaptic weight information with a density comparable to the density in electronic chips. It could be further increased if holographic encoding of some form was be used. Assuming a conservative estimate of encoding a single weight parameter on $10\,\mu$m$^2$ of surface, the size of a weight bank encoding the full BERT language model with 110 million parameters would require a surface of only 11~cm$^2$.  

\subsection{Speed} 

Speed of a neural network inference can be measured in several ways. While the number of operations per second is a valid measure of computational power, probably more important ones from the practical point of view are latency and performance density, which is the number of operations per second per area~\cite{Xu_EdgeInference}.  In terms of latency, all-optical networks can certainly outperform electronic and optoelectronic networks in most applications, since apart from input generation and output detection, they require only propagation of light across the network layers at the speed of light. For a centimeter-sized system, this results in latency of the order of picoseconds, which is many orders of magnitude lower than millisecond latency typical for electronics~\cite{Jouppi_InDatacenter}. Performance density of all-optical networks can be estimated by considering the number of synaptic weights multiplications per second, taking into account the 10 $\mu$m$^2$ weight footprint as estimated above and 10 GHz inference rate corresponding to commercial optical modulators. The resulting performance density of the order of 10$^6$ GOP s$^{-1}$ mm$^{-2}$ is three orders of magnitude higher than in state-of-the-art electronics~\cite{Xu_EdgeInference}.

\subsection{Strength of nonlinearity} 

An efficient all-optical neural network requires strong optical nonlinearity. This nonlinearity has to be characterized by a fast response time, ideally in the GHz to THz range, since the optical pulse energy required for realizing an operation scales linearly with its duration at the same light intensity. A range of optical materials have been considered for this purpose~\cite{hill2002all,rosenbluth2009high,Feldmann_AllOpticalSpikingNetwork,guo2022femtojoule}.
 In particular, semiconductor materials possess characteristics that make them good candidates for nonlinear  elements of artificial 
 neurons~\cite{hill2002all,rosenbluth2009high,Brunner_LowThreshold}.  

In this context, microcavity exciton-polaritons~\cite{Opala_NeuromorphicComputing,Matuszewski_EnergyEfficient,Ballarini_Neuromorphic,Mirek_Neuromorphic} are a particularly promising alternative. These quasiparticles are half-light half-matter excitations existing in semiconductors, which induce optical nonlinearity orders of magnitude stronger than in standard semiconductor materials~\cite{Matuszewski_EnergyEfficient}. They can operate at room temperature~\cite{Fieramosca_perovskites,Su_perovskites,Lagoudakis_RTOrganicTransistor} and have response times of hundreds of femtoseconds to nanoseconds~\cite{bobrovska2018dynamical}. Moreover, the nonlinearity of polaritons is significantly enhanced in two-dimensional materials~\cite{datta2022highly}, in the case of trion-polaritons~\cite{emmanuele2020highly}, or Rydberg polaritons~\cite{gu2021enhanced}.

\subsection{Activation functions} 

Electronic implementations of neural networks make it possible to realize virtually any nonlinear activation function at a very low energy cost. In all-optical networks, one does not have such a flexibility and usually has to deal with a nonlinear response of the system that is either fixed or exhibits some limited tunability. Moreover,  it is usually not possible to realize the activation function that is optimal for a particular network model. However, as is well known in the field of machine learning, the particular form of activation function often has only limited impact on system performance. On the other hand, the use of real and imaginary parts of the complex optical field amplitude may lead to certain improvements in accuracy~\cite{CVNN}.
\begin{figure}
    \centering
    \includegraphics[width=0.9\columnwidth]{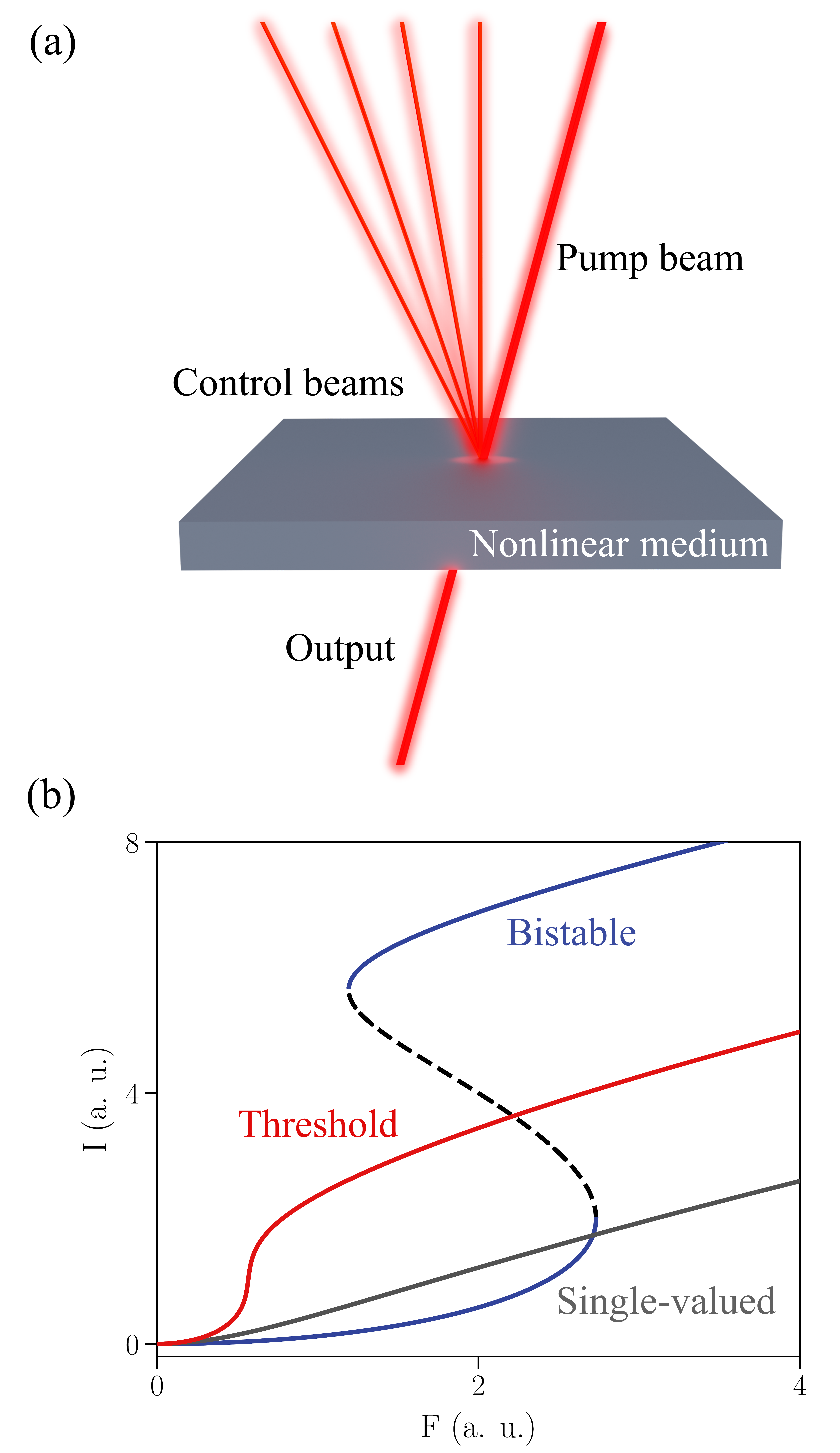}
    \caption{A possible realization of an optical neuron (a). A nonlinear system is tuned close to the bistability threshold which results in sigmoid-like response to the total intensity of control beams. The transmittance of the strong pump beam depends on the nonlinear index change induced by weak control beams. As a result, both nonlinear activation and signal amplification can be realized.
(b) Schematic examples of output intensity $I$, as a function of total incident amplitude $F$ in the bistable, threshold, and single-valued cases.
}
    \label{fig:nonlinearity}
\end{figure}
We checked the impact of the type of activation function and the complex nature of light field by considering a simple example of a feedforward neural network with one hidden layer performing the MNIST handwritten digit recognition task. We consider the following real and complex activation functions
 \begin{align} 
 f_j(x) &= \begin{cases}
 \mbox{ReLU}(x) \\
 \frac{1}{1+\ee^{-x}}
 \end{cases} \\
 f_j(z) &= \begin{cases}
 |z| \\
 \frac{1}{1+\ee^{-|z|}}
 \end{cases} \label{eq:sigmoid}
 \end{align}
The first two functions are standard activation functions used in machine learning. Note that in the case of functions in Eq.~(\ref{eq:sigmoid}), which take complex arguments, both the inputs $x_i$ and the weights $w_{ij}$ in Eq.~(\ref{eq:neuron}) can be complex-valued, which reflects the amplitude and phase of the optical field. The first function  in Eq.~(\ref{eq:sigmoid}) is one of the simplest nonlinear functions that takes advantage of the complexity of variables.  To justify the form of the second function in Eq.~(\ref{eq:sigmoid}), we consider a simple optical setup shown in Fig.~\ref{fig:nonlinearity}. This setup is based on the nonlinear refractive index change induced by both the optical control beams and the signal beam, which is transmitted through the nonlinear medium. The optical nonlinearity can be enhanced by enclosing the medium in a microcavity and achieving strong light-matter coupling~\cite{Matuszewski_EnergyEfficient}. At the point of the optical bistability threshold Fig.~\ref{fig:nonlinearity} (b), (red line), the  dependence of the transmitted signal intensity on the incident light amplitude is strongly nonlinear. It exhibits an S shape analogous to the sigmoid activation function known from machine learning models. We assume that control beams and the input beam are not coherent (eg.~formed by different lasers), which allows us to discard the effects of interference. On the other hand, control beams are assumed to be coherent with each other. This assumption is natural if coherent light is used in the linear vector-matrix multiplication setup, which precedes the nonlinear activation stage~\cite{goodman1978fully,Soljacic_DeepLearning}. The phase of the transmitted beam is therefore not related to the phases of control beams, but its intensity is strongly modulated by the intensity of the superposition of control beams. Here, we assume a simplified, complex sigmoid dependence of the transmitted light intensity at the threshold
\begin{equation}
    I_{\rm out} \sim I_{\rm pump} \frac{1}{1+{\rm e}^{-|z|}}
\end{equation}
where $z=\sum A_i$ is the sum of complex amplitudes of all of the control beams corresponding to this nonlinear node. These beams can be treated as synaptic inputs to the nonlinear node. Therefore, we consider the situation where the intensity of the pump beam is tuned to the middle of the sigmoidal dependence near the optical bistability threshold, see red line in Fig.~\ref{fig:nonlinearity} (b).

\begin{figure}
    \centering
    \includegraphics[width=\columnwidth]{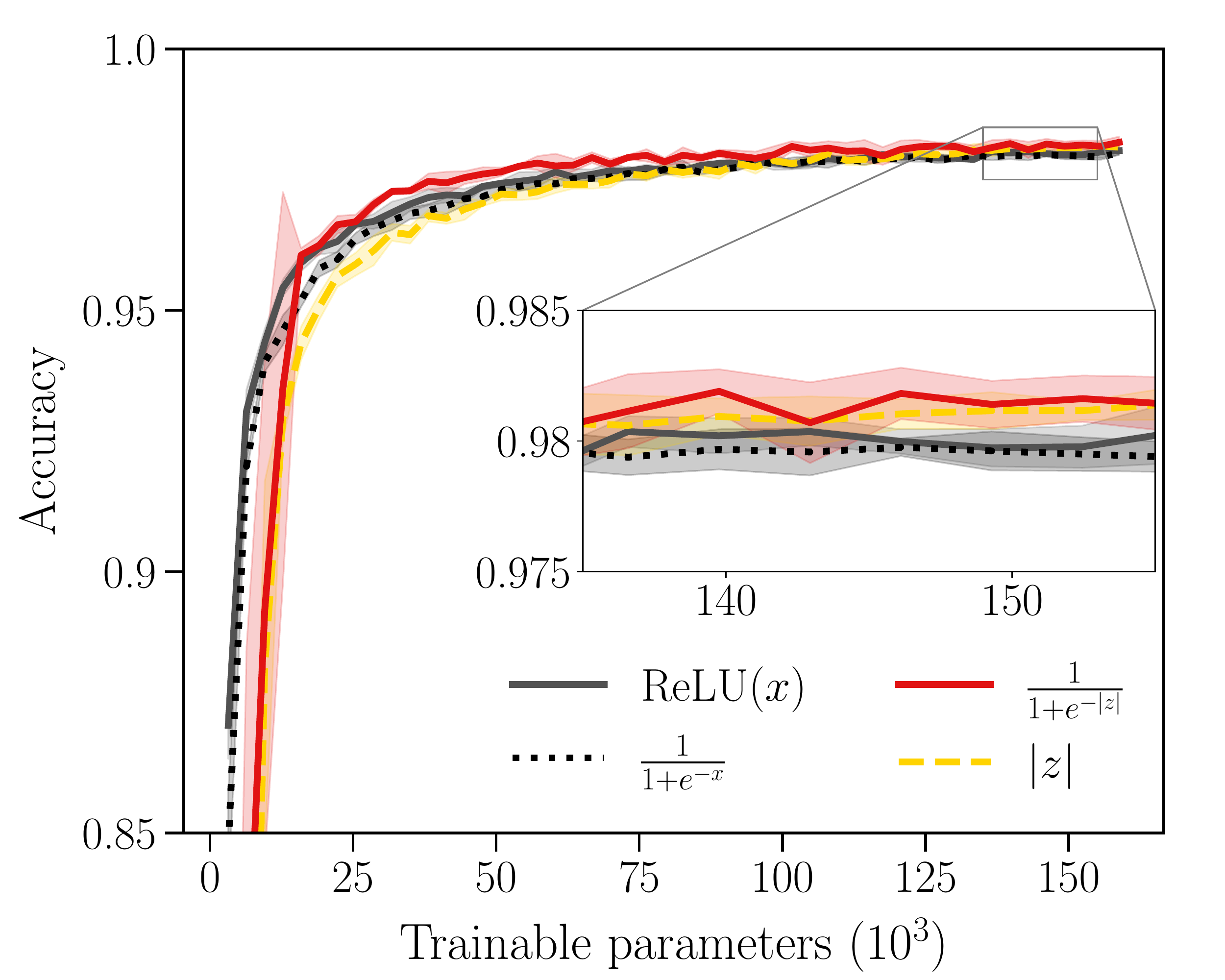}    \caption{Accuracy of handwritten digit recognition for fully connected complex- and real-valued networks with 1 hidden layer as a function of the number of trainable parameters. Different activation functions correspond to electronic, optoelectronic and all-optical neurons. Results after 150 epochs of training are shown. Shaded regions correspond to the estimated uncertainty based on multiple training iterations. Neurons in the hidden layer with complex activation functions were modeled without biases due to the possible difficulty of their optical implementation.} 
    \label{fig:100nodes}
\end{figure}

In Fig.~\ref{fig:100nodes} we present the estimated accuracy of fully connected complex-valued and real-valued neural network models with different nonlinear activation functions. Two conclusions can be drawn from these results. First, complex-valued networks can perform slightly better than real-valued networks with the corresponding activation functions and the same number of parameters. This is the case even if biases are not used in the complex valued networks, which simplifies the implementation with optics. Second, the particular form of activation functions can have some influence on the accuracy, but there is no substantial difference between "optimal" functions such as the ReLU function and the sigmoid. In particular, the physically relevant complex sigmoid activation controlled by complex-valued inputs gives optimal results.

\subsection{Precision} 

Limited precision of analog systems is a potential obstacle for applications. In the context of neural networks, it is known that low precision can be good enough to perform machine learning tasks with very high accuracy as long as it is kept above a certain, task-dependent level. Examples include quantized and binarized neural networks~\cite{Bengio_Binarized,Rastegari}. In the context of analog computing, there are examples where 2-6 bit precision is sufficient to achieve accuracy close to an optimal one~\cite{Rastegari,Merolla,mishra2017wrpn}. It appears that the required precision of computations strongly depends on the task to be solved and must be considered on a case-by-case basis.

\subsection{Fabrication errors} 

The influence of fabrication variability can strongly impact system performance. Ideally, robustness would mean that a neural network model trained {\it in silico} can perform equally well in a physical system where the parameters are not fully controllable. This can be achieved by reducing device variability, but it is not always possible. However, additional post-processing correction methods or tunable "control knobs" may be used to adjust the system. For example, in the scheme shown in Fig.~\ref{fig:nonlinearity}, such knobs is the phase of the pump beam, and the weights of the linear vector-matrix multiplication, which can correct for the variability of the nonlinear response of the sample. Another method is to fine-tune a pretrained model taking into account the imperfections existing in a particular device, or specific training methods that directly take into account the response of the physical system~\cite{McMahon_PAT,spall2022hybrid}. If the system is to be used many times in the inference phase, performing such procedures once for each device may be reasonable, even if they are lengthy or expensive.

\begin{figure}
    \centering
    \includegraphics[width=\columnwidth]{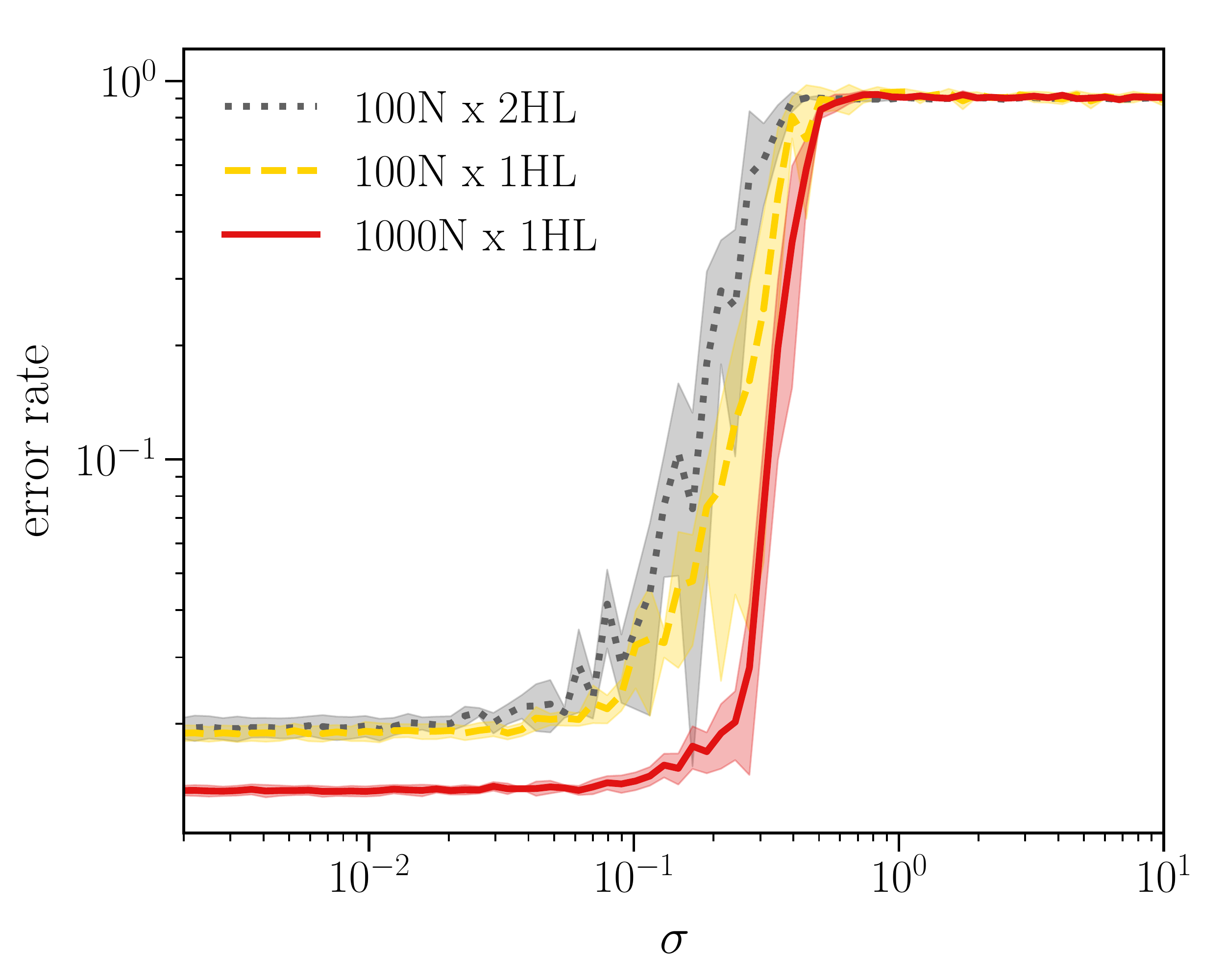}
    \caption{Influence of imperfections on the performance of optical neural networks. A static disorder of relative amplitude $\sigma$ is applied to each neuron as described in Eq.~(\ref{eq:sigma}). Networks with one and two hidden layers (HL) are considered, with 100 or 1000 neurons (N) in each hidden layer. Error rate is defined here as $1 - \eta$, where $\eta$  is the accuracy of the model.}
    \label{fig:disorder}
\end{figure}

Apart from these correction methods, neural networks are characterized by an intrinsic robustness to imperfections. To investigate the robustness of optical networks, we analyze the accuracy of the neural network used for the MNIST dataset classification in the case when an additional disorder is introduced to individual neurons. In Fig.~\ref{fig:disorder}, we show the accuracy in function of disorder strength, which is perturbing the response of hidden neurons only in the inference phase, according to
\begin{equation}\label{eq:sigma}
f_j(z)=\frac{a_j}{b_j+c_je^{-|z|}}
\end{equation}
where $a_j,b_j,c_j$ are parameters chosen individually for each neuron from the distributions described by a Gaussian probability density centered around unity i.e. $p(x)=\frac{1}{\sqrt{2\pi\sigma^2}}e^{-\frac{1}{2}\left(\frac{x-1}{\sigma}\right)^2}$, where $p(x)$ is the probability density of $a_j,b_j,c_j$ taking the value of $x$ and $\sigma$ is the factor describing the width of the distribution and thus the strength of disorder. The results shown in  Fig.~\ref{fig:disorder} indicate that up to a certain disorder strength, the accuracy of the network inference does not suffer significantly. Robustness certainly increases with the number of neurons, but is decreases in with the number of layers, which can be interpreted as propagation of errors. This shows that even in the case when correction of device imperfections is not possible, reducing the disorder below a certain level may be sufficient.

\subsection{Quantum Noise}  

One of the benefits of all-optical computing is the absence of thermal noise at the intermediate layers of computation, which is inevitably present in electronic elements whenever optoelectronic conversion occurs. The fundamental limit of energy efficiency of all-optical computing is related to quantum noise, or shot noise, which becomes significant near the single photon level. In the context of optoelectronic networks, it was shown both theoretically~\cite{Englund2019} and experimentally~\cite{McMahon_SinglePhoton} that vector-matrix multiplication  operations in neural networks can be performed, even below the single photon per operation level. The reason for this surprising result is that when many such operations contribute to the result of the weighted summation in a single neuron as in Eq.~(\ref{eq:neuron}), the signal to noise ratio of the sum is approximately a factor of $\sqrt{N}$ higher than the ratio for individual elements of the sum. This property can greatly increase the energy efficiency of neural networks if the number of neuron inputs is large, which is the case in many practical neural network models.

We analyze to what extent all-optical neural networks, where information is encoded with coherent light amplitudes, can benefit from a similar quantum noise reduction. We assume that the weighted input of an optical neuron $j$, $w_{ij} x_i$ in Eq.~(\ref{eq:neuron}) is encoded by a coherent optical laser beam of amplitude proportional to $w_{ij} x_i$. 
Recall that both the weights $w_{ij}$ and the inputs $x_i$ are complex-valued. A superposition of such beams results in optical amplitude proportional to the weighted summation as in Eq.~(\ref{eq:neuron}) for a given neuron $j$. In the following, focus on a particular neuron and drop index $j$ for convenience.

In our quantum treatment, weighted inputs are optical laser pulses represented by coherent photon states $|\alpha_i\rangle$ such that $\alpha_i = \beta w_i x_i$, where $\beta$ is a factor that relates the coherent state amplitude to the amplitude of the neuron input. For a given neural network model, it can be chosen arbitrarily, with higher values of $\beta$ resulting in stronger light intensities and a higher signal-to-noise ratio. The approximation of treating inputs as coherent states may not be correct when one is dealing with input states that are quantum themselves, for example, when they have been affected by a strong single-photon nonlinearity in the previous computation layer. In the following, we will exclude this possibility, which is consistent with the fact that the nonlinearity of optical materials does not allow to achieve such a strong single-photon nonlinearity except of very specific configurations~\cite{arakawa2020progress,munoz2017quantum}.

    For convenience, we denote weighted inputs with $a_i=w_i x_i$. Thus, we take  input states of the neuron as $|\alpha_i \rangle = \beta a_i$ where $i=1\ldots N$, and assume that the output state is approximately a coherent state. The weighted sum of inputs, i.e., the state of light in the spatial and temporal mode corresponding to the neuron, is a superposition of $N$ input states, i.e., $|\alpha \rangle=\sum_{i=1}^N |\alpha_i \rangle$, and it is also a coherent state with $\alpha=\sum_i \alpha_i$. We neglect phase factors such as $\ee^{i(kr - \omega t)}$ since we can select the basis for input coherent states in such a way that they are eliminated. 

We can now determine all the quantum properties of light, in particular its intensity and fluctuations. To determine the fluctuations it is convenient to use quadratures $\hat{X}_1$ and $\hat{X}_2$, with $\alpha=\langle \alpha| \hat{X}_1|\alpha\rangle + i \langle \alpha| \hat{X}_2|\alpha\rangle$.  In our neural network model, we simulate quantum noise by introducing $a_i'=a_i+\delta a_i$ with $\delta a_i$ being random variables reproducing quantum shot noise, with appropriate statistics. It is clear that the expectation value of $a_i'$ should be equal to  $\bar{a'_i}=a_i=\alpha_i/\beta=(\langle \alpha_i| \hat{X}_1|\alpha_i\rangle + i \langle \alpha_i| \hat{X}_2|\alpha_i\rangle)/\beta$. On a similar basis, the fluctuations $\delta a_i$ will also scale proportionally to $1/\beta$, so finally we get $a'_i=a_i+\delta a_i$, where $\delta a_i$ is a complex Gaussian noise with variance
\begin{align}
    (\Delta \Re (\delta a_i))^2 &=((\Delta \hat{X}_1)^2)/\beta^2 = 1/4\beta^2,\\
    (\Delta \Im (\delta a_i))^2 & =((\Delta \hat{X}_2)^2)/\beta^2 = 1/4\beta^2.
\end{align}
This defines the statistical properties of $a_i'$ which we use in numerical simulations. At the same time, we can determine the average energy of input light pulses from the formula $E_i=\hbar \omega|\alpha_i|^2$, which scales proportionally to $|\beta|^2$.

\begin{figure}
    \centering
    \includegraphics[width=\columnwidth]{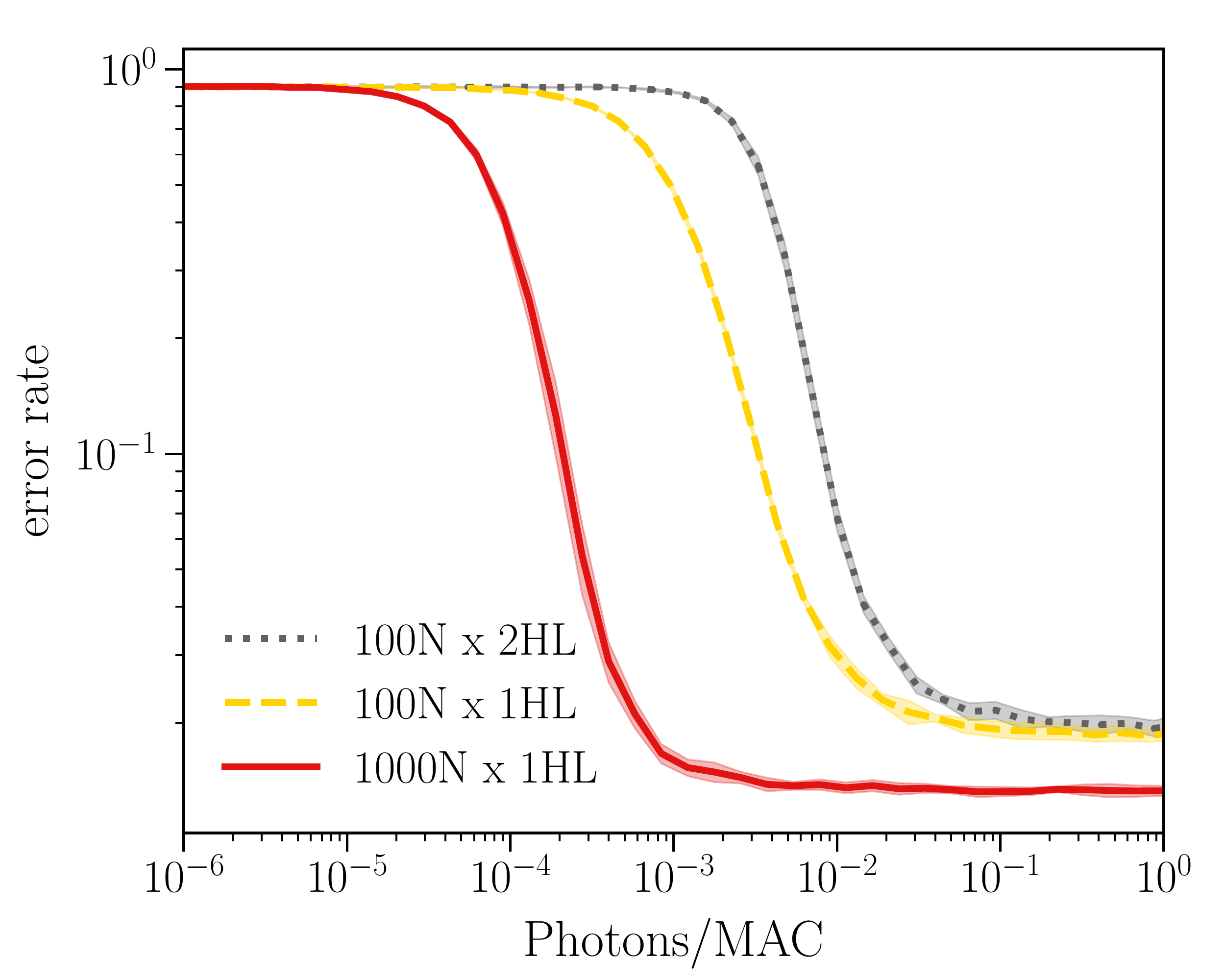}
    \caption{Error rate of neural networks as a function of the number of photons per operation, with quantum noise included. Results are shown for networks with a single hidden layer (1HL) and two hidden layers (2HL). It was assumed that the $\beta$ factor relating coherent state amplitude to the amplitude of the neuron input is the same in all layers.}
    \label{fig:noise}
\end{figure}

We present the result of simulations of optical networks with quantum noise included in Fig.~\ref{fig:noise}. As in the optoelectronic case~\cite{Englund2019}, we find that the error rate of predictions can remain low even in the case when the number of photons per operation is lower than unity.  In the optical range, this corresponds to hundreds of zeptojoules per operation. As a result, we may expect that quantum noise will not be a limiting factor up to this level of energy efficiency.

\subsection{Signal degradation and network depth} 

While optical signal regeneration and amplification is possible~\cite{hill2002all,rosenbluth2009high}, signal decay and degradation is one of the most important challenges for all-optical systems, especially in the case of multi-layer networks. Although cascadability of optical neurons has been achieved~\cite{tait2019silicon}, full cascadability in a large-scale system may be difficult to realize in practice. Moreover, elements of the optical setup necessary for light beam steering may lead to significant losses~\cite{mccormick1993six}. In the case where full regeneration is not viable, or signal distortion at each layer is significant, these factors will limit the possible number of network layers. Since the most successful applications of machine learning are based on deep networks, with up to thousands of layers in networks such as ResNet, this is an obstacle that could limit the practical use of all-optical networks. 

Recent results in the field of machine learning suggest that the number of neural networks layers can often be greatly reduced without the loss of accuracy, if model designs are appropriately modified. Examples include shallow networks for speech and image recognition~\cite{Ba_Deep,Urban_Deep}, non-deep networks achieving state-of-the-art results with a reduced number of layers~\cite{Goyal_NonDeep} and shallow transformer networks for language models, that successfully compete with recurrent neural networks~\cite{devlin-etal-2019-bert}. In many of these cases, one or two hidden layers are enough to obtain high accuracy of predictions. Some authors suggested that in the case of fully connected or convolutional networks, making networks deeper beyond a relatively shallow level does not improve accuracy~\cite{Winkler_Deep,thomas2017two}. These observations are aligned with the arguments considering models of physical systems, which are usually described by  Hamiltonians that are low-order polynomials~\cite{lin2017does}.

\subsection{Neural network architectures}

All-optical networks have limited flexibility of possible architectures. This concerns also the structure of computations. Some neural network architectures are easier to implement than others. For example, it may be straightforward to design an optical feed-forward neural network with scalar nonlinear activation functions, but more complex models require vectorial nonlinear operations. It is known that vector-matrix multiplications can be implemented optically as long as the vector component is encoded optically, while the matrix component is encoded in the material part of the device~\cite{goodman1978fully}. The same is true for convolutions~\cite{Chang_CNN,Feldmann_parallel,Xu_TOPS}. However, it is not known how to implement all-optically some other nonlinear transformations that are important for neural network models. These include vector-matrix multiplications and softmax activations, which are key components of attention layers~\cite{vaswani2017attention,McMahon_Transformers}, where both the vector and the matrix are to be encoded optically. It is important to either find a way to realize these functions all-optically, or to determine alternative models that do not require these functions but are able to perform the same tasks with comparable accuracy.

\subsection{Constant radiance theorem} 

Constant radiance theorem imposes a fundamental limitation on the geometry and optical energy required for performing computations with light~\cite{Goodman_FanIn}. This theorem states that in the case of linear propagation of light the generalized \'etendue, which measures the spread of light in real and momentum space, remains constant. In the case of neural networks, this condition imposes a limit on the optical energy per neuron, which inversely scales with the number of neurons. In particular, if the number of neurons in a hidden layer is much higher than in the input layer, the energy per neuron in the hidden layer is allowed to be much lower than the average energy of optical inputs. As a result, if the condition stated in Sec.~\ref{sec:alloptical} is fulfilled, that is $N_{\rm hidden}\gg N_{\rm input} + N_{\rm output}$,  high energy efficiency of operations in the hidden layer is not excluded by constant radiance theorem. On the other hand, in the case of small neural networks or networks in which this condition is not fulfilled, constant radiance theorem will impose a limit on the achievable energy efficiency. For all-optical networks, this has to be considered as an important factor in system design.

\section{\label{sec:conclusions} Conclusions} 

In conclusion, under certain plausible assumptions about the limitations of electronics, we showed that all-optical neural networks can find an important role in the applications of machine learning. It is estimated that all-optical devices could outperform both electronic and optoelectronic devices by orders of magnitude in energy efficiency in the case of inference in large neural network models. This estimate takes into account all the components of the complete system, including the cost of memory access and data acquisition from remote resources. All-optical networks are predicted to give the biggest advantage in the case of generative models, where the cost of data acquisition and memory access is reduced due to the possibility to reuse input data.

On the other hand, it is clear that there are still important issues that need to be solved before all-optical networks become practical. These include scalability of optical neurons, signal decay and distortion, strength of nonlinearity, and non-universal character of optical computing. To overcome these obstacles, studies on both the physical implementations of optical networks and on accommodating neural network models to the capabilities of optical systems may be necessary. It is likely that an interdisciplinary approach will be the key to successful implementations.

\acknowledgments

 We acknowledge support from the National Science Center, Poland grants 2019/35/N/ST3/01379, 2020/37/B/ST3/01657 and 2021/43/B/ST3/00752.

\bibliography{bibliography}

\end{document}